\documentclass[letterpaper,twocolumn,10pt]{article}
\usepackage{usenix-2020-09}

\usepackage{amsmath}
\usepackage{cleveref}
\usepackage{graphicx}
\usepackage{makecell}
\usepackage{hyperref}

\usepackage{authblk}

\usepackage[skip=5pt]{caption}
\usepackage[skip=0pt]{subcaption}
\usepackage{enumitem}
\setlist{noitemsep}

\usepackage{etoolbox}
\makeatletter
\patchcmd{\maketitle}
	{\@maketitle}
	{\@maketitle\vspace{-4em}}
	{}
	{}
\makeatother

\makeatletter
\renewcommand{\paragraph}{%
  \@startsection{paragraph}{4}%
  {\z@}{0.5ex \@plus 1ex \@minus .2ex}{-1em}%
  {\normalfont\normalsize\bfseries}%
}
\makeatother

\usepackage{xcolor}
\usepackage{pifont}
\newcommand{\yes}{{\color[rgb]{0,0.5,0}\ding{51}}}
\newcommand{\no}{{\color[rgb]{0.5,0,0}\ding{55}}}
\newcommand{\half}{{\color[rgb]{0.5,0.5,0}½}}

\usepackage{listings}
\definecolor{mygreen}{rgb}{0,0.6,0}
\definecolor{mygray}{rgb}{0.5,0.5,0.5}
\definecolor{mymauve}{rgb}{0.58,0,0.82}
\definecolor{darkgray}{rgb}{.4,.4,.4}
\definecolor{purple}{rgb}{0.65, 0.12, 0.82}
\lstdefinelanguage{TypeScript}{
keywords={typeof, new, true, false, catch, function, return, null, catch, switch, var, if, in, while, do, else, case, break},
keywordstyle=\color{blue}\bfseries,
ndkeywords={class, export, boolean, throw, implements, import, this, constructor, extends, protected, private, readonly, super},
ndkeywordstyle=\color{black}\bfseries,
identifierstyle=\color{black},
sensitive=false,
comment=[l]{//},
morecomment=[s]{/*}{*/},
commentstyle=\color{purple}\ttfamily,
stringstyle=\color{red}\ttfamily,
morestring=[b]',
morestring=[b]",
}
\begin{document}

\date{}

\title{\Large \bf Collabs: A Flexible and Performant CRDT Collaboration Framework}
\author[1]{Matthew Weidner*}
\author[1]{Huairui Qi}
\author[2]{Maxime Kjaer}
\author[1]{Ria Pradeep}
\author[3]{Benito Geordie}
\author[1]{Yicheng Zhang}
\author[4]{Gregory Schare}
\author[1]{Xuan Tang}
\author[1]{Sicheng Xing}
\author[1]{Heather Miller*}
\affil[1]{Carnegie Mellon University}
\affil[2]{EPFL IC}
\affil[3]{Rice University}
\affil[4]{Columbia University}

\renewcommand\Authands{ and }

\maketitle

\begin{abstract}
A collaboration framework is a distributed system that serves as the data layer for a collaborative app. Conflict-free Replicated Data Types (CRDTs) are a promising theoretical technique for implementing collaboration frameworks. However, existing frameworks are inflexible: they are often one-off implementations of research papers or only permit a restricted set of CRDT semantics, and they do not allow app-specific optimizations. Until now, there was no general framework that lets programmers mix, match, and modify CRDTs.

We solve this with Collabs, a CRDT-based collaboration framework that lets programmers implement their own CRDTs, either from-scratch or by composing existing building blocks. Collabs prioritizes both semantic flexibility and performance flexibility: it allows arbitrary app-specific CRDT behaviors and optimizations, while still providing strong eventual consistency. We demonstrate Collabs's capabilities and programming model with example apps and CRDT implementations. We then show that a collaborative rich-text editor using Collabs's built-in CRDTs can scale to over 100 simultaneous users, unlike existing CRDT frameworks and Google Docs. Collabs also has lower end-to-end latency and server CPU usage than a popular Operational Transformation framework, with acceptable CRDT metadata overhead.
\end{abstract}

\vspace{-3mm}

\begingroup\def\thefootnote{*}\footnotetext{Corresponding authors: Matthew Weidner \href{mailto:maweidne@andrew.cmu.edu}{maweidne@andrew.cmu.edu}, Heather Miller \href{mailto:heather.miller@cs.cmu.edu}{heather.miller@cs.cmu.edu}.}\endgroup

\section{Introduction}
\label{sec:intro}
\vspace{-2mm}

In a collaborative app, users expect to see their own operations immediately, without waiting for a round-trip to a central server. Local-first apps \cite{local_switch} take this further and allow users to perform local operations even when they are not connected to a central server, whether due to offline work, decentralization, or git-style ``forks'' of documents \cite{upwelling}. Local operations may make users temporarily see different states; later, when they synchronize with each other, they will converge to a state that combines all of their operations.

In distributed systems terms, local-first apps are Available and Partition-tolerant, but only (Strongly) Eventually Consistent \cite{strong_eventual_consistency_def}. This is the AP side of CAP \cite{cap}.

Designing local-first apps, and traditional collaborative apps, poses a difficult challenge: we must decide what the state should be after multiple users perform operations concurrently. The ``right'' answer is not always clear, and it depends on the application's specific semantics and what users expect to happen \cite{peritext, ignat_rich_text, update_wins, for_each, json_crdt}.

\begin{figure}
    \centering
    \includegraphics[width=0.95\columnwidth]{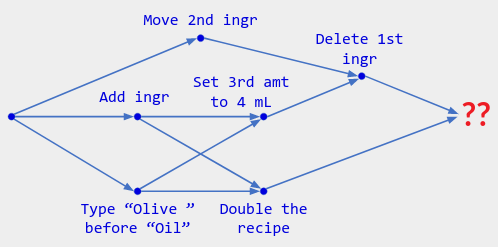}
    \caption{A possible operation history for a collaborative recipe editor.}
    \label{fig:history_dag}
\end{figure}

For example, \Cref{fig:history_dag} shows a possible operation history for a collaborative recipe editor. Time proceeds to the right, while arrows indicate causal dependencies---operations that were aware of each other---so that concurrent operations are in parallel. Given this history, what should the editor's state be? How can we compute that state efficiently?

Conflict-free Replicated Data Types (CRDTs) \cite{crdt_survey_2011} provide a potential solution to this problem. CRDTs are algorithms that compute an app's state from a history like \Cref{fig:history_dag}. Specifically, they tell you how to update the state in specific practical situations:
\begin{description}
    \item[Op-based CRDTs] update the state to reflect a single new operation. This is useful for live collaboration, when you are receiving live updates from remote collaborators.
    \item[State-based CRDTs] ``merge'' two operation histories, updating the state to reflect their combined changes. This is useful for peer-to-peer synchronization and for combining multiple versions of a document, e.g., merging the local disk version with a server's latest version.
\end{description}

Open-source CRDT implementations exist for a few specific collaborative apps like rich-text editing \cite{peritext}, as well as for basic data structures like sets and maps \cite{crdt_survey_2011}.

However, for the vast majority of apps, there is no published CRDT implementation. Thus we need a collaboration framework that lets programmers implement new CRDTs for their own apps. A good framework should have two key features:
\vspace{-2mm}
\begin{description}
    \item[Semantic flexibility] Programmers can choose what the state should be after multiple users perform operations concurrently. These choices can be nuanced and app-specific. For example, suppose one user deletes an ingredient while another is still editing it concurrently. The programmer should get to choose whether the ingredient remains deleted (\emph{delete-wins}), is revived by the concurrent edits (\emph{update-wins}) \cite{update_wins}, or is archived but can be restored later (\emph{archiving}).
    \item[Performance flexibility] Programmers can optimize CRDT implementations at a low level. In critical cases, programmers can choose the binary content of network messages and their exact message-processing algorithms. 
    Thus programmers can fine-tune network, disk, CPU, and memory performance for their own applications.
\end{description}
\vspace{-1mm}

Unfortunately, existing collaboration frameworks offer neither of these features. Instead, they dictate the allowed behaviors to apps: either they provide ``one-size-fits all'' algorithms \cite{automerge, yjs, sharedb}, or they derive CRDTs with unclear semantics from sequential code \cite{mrdts, ecros, katara}.
Often, the results of concurrent operations are not documented, or they depend on unpredictable runtime events. Performance optimizations are impossible without changing a framework's source code.

\vspace{-3mm}

\subsection{Collabs}
\label{sec:intro_collabs}
\vspace{-2mm}

This paper introduces Collabs, a collaboration framework that prioritizes both semantic flexibility and performance flexibility. Specifically, Collabs is a framework for hybrid op-based/state-based CRDTs that work efficiently in both of the practical situations described above.

Our insight is that any CRDT framework with a small, rigid set of techniques cannot satisfy all collaborative apps. Instead, collaborative app programmers need a library of building blocks that let them mix, match, and modify CRDTs. Thus Collabs lets programmers implement arbitrary CRDTs, allowing them to customize the semantics of apps like \Cref{fig:history_dag}'s recipe editor. For example, all three semantics for deleted ingredients are possible in Collabs: delete-wins, update-wins, and archiving. We walk through a collaborative recipe editor and further semantic choices in \Cref{sec:flexibility}.

In particular, programmers can create new CRDTs using composition techniques. A programmer who composes existing CRDTs still needs to evaluate whether the resulting semantics are appropriate for their application, but they do not need to worry about the difficult parts of CRDTs: correctness guarantees (e.g., ensuring strong eventual consistency), efficient message encodings, and state-based merge functions are all inherited from the components. We demonstrate composition by implementing two cutting-edge CRDT algorithms, the Peritext rich-text CRDT \cite{peritext} and a list with a move operation \cite{list_with_move}, in a composed way. \Cref{app:composed_algs} walks through the latter implementation.

We evaluate Collabs by comparing it to several existing collaboration frameworks, including the local-first CRDT libraries Yjs \cite{yjs} and Automerge \cite{automerge}. \Cref{sec:eval_cap} compares the capabilities that different frameworks offer. These affect how easily programmers can develop collaborative apps on top of the frameworks. Collabs is the only CRDT library that supports semantic flexibility and performance flexibility, and the only collaboration framework that supports encapsulated custom data models (\Cref{sec:composition}).

Finally, \Cref{sec:eval_perf} evaluates Collabs's performance on a collaborative rich-text editing benchmark. We show that Collabs has state-of-the-art performance on this popular but difficult application: using its built-in CRDTs, our rich-text editor can support 112 simultaneous users typing at realistic speeds, compared to at most 16 for Google Docs and Automerge, and 96 for Yjs (\Cref{tab:max_users}). While ShareDB---a popular Operational Transformation framework \cite{sharedb}---can also support 112 users, Collabs has noticeably lower end-to-end latency and server CPU usage (\Cref{tab:sharedb_vs_collabs}), with acceptable CRDT metadata overhead (\Cref{tab:metadata}).


Collabs is written in TypeScript and published on npm.\footnote{\url{https://www.npmjs.com/package/@collabs/collabs}} It is open source on GitHub,\footnote{\url{https://github.com/composablesys/collabs}} fully documented,\footnote{\url{https://collabs.readthedocs.io/}} and comes with several demo collaborative apps. We also provide practical tools like a testing server and React hooks.

In the future, we hope that Collabs can serve as a ``CRDT laboratory'': researchers can implement and publish new algorithms as simple extensions. That would let app programmers use new CRDT algorithms immediately, mix-and-match approaches, and even optimize existing approaches. Extensions can be used in concert with the rest of Collabs's CRDTs and practical plugins, instead of requiring a new collaboration framework per research paper.


\vspace{-3mm}

\section{System Model}
\label{sec:model}
\vspace{-3mm}

As a prerequisite, we describe Collabs's system model. It is a hybrid of system models for op- and state-based CRDTs \cite{crdt_survey_2011}.

\paragraph{Documents, Users, and Replicas}
A collaborative app interacts with Collabs through a \emph{document}, which is a collection of CRDTs that are shared together. For example, a document could be a single shared recipe, a collaborative rich-text document, or a shared whiteboard. A document is shared between a group of \emph{users}, each of whom has read or read-write access to the entire document. An app may use multiple documents, but in the presentation below, we always assume just one.

To access a document, a user creates a \emph{replica} on their local device. Each replica functions as a node in a distributed system. We blur the distinction between replicas and users, but technically, there can be multiple replicas per user (e.g., if a user opens the same document in multiple browser tabs).

\paragraph{Op-based CRDT Usage}
At any time, a user with read-write access to a document may perform an \emph{operation} on their replica of that document (i.e., on one of its component CRDTs). The replica immediately updates its state to reflect that operation. It also generates a \emph{message} that the user may broadcast to other replicas of the document. Any other user may deliver that message to their own replica of the document. There are no requirements on the network used to broadcast messages; in particular, it may be decentralized, and it may suffer unbounded network latencies.

When a replica receives a message, it does not necessarily update its own state immediately. Instead, the Collabs runtime delays or ignores messages in order to guarantee \emph{exactly-once causal order delivery} to the internal CRDTs: each message is delivered exactly once; and a message $m$ is not delivered until after all messages that are causally prior to $m$ (i.e., happened-before $m$) \cite{lamport_causal}. Exactly-once causal order delivery is a standard assumption of op-based CRDT algorithms, and it also ensures \emph{causal consistency}: for example, if one user adds an ingredient to a recipe and then writes instructions for it, all users will see the ingredient before the instructions \cite{causal_plus_consistency}. Internally, causal order delivery is implemented using causally-maximal vector clock entries, which omit entries from (potentially numerous) inactive replicas \cite{fidge, mattern}.

A Collabs document guarantees \emph{strong convergence}: two replicas that have received the same messages will be in equivalent states. For example, if two users make concurrent offline edits to their own replicas of a Collabs shared recipe, then once they come online and exchange messages, they will converge to the same state. (The contents of this state are determined by the app's chosen CRDTs, as we illustrate in the next section.) It follows that, in any network that eventually delivers messages at-least-once, a Collabs document satisfies \emph{strong eventual consistency}: two replicas who stop performing operations will eventually be in equivalent states \cite{strong_eventual_consistency_def}.

\paragraph{State-based CRDT Usage}
So far, we have described an op-based CRDT system model. Collabs also supports a state-based CRDT system model. At any time, the app may ask its replica for a \emph{saved state}. This is a byte array that incorporates all operations delivered to the replica so far.\footnote{Note that a saved state may lossily encode the original operations, e.g., by omitting text that has since been deleted.}

Another replica---possibly another user's---may \emph{load} this saved state at any time. Collabs guarantees that loading a saved state is equivalent to delivering all of the saved state's incorporated operations, skipping duplicates. Thus loading performs a state-based CRDT merge, and the document has equivalent op- and state-based semantics. Of course, saving and loading can also be used for ordinary disk storage.

\vspace{-4mm}

\section{Semantic Flexibility}
\label{sec:flexibility}
\vspace{-3mm}

Recall that Collabs prioritizes semantic flexibility: programmers can choose what the state should be after multiple users perform operations concurrently. To illustrate Collabs's semantic flexibility, let us walk through a collaborative recipe editor app built on top of Collabs.\footnote{
Source code is available at \url{https://github.com/composablesys/collabs/tree/master/demos/apps/recipe-editor}.
}
This app lets friends or family members share a recipe and edit it on multiple devices, including while offline. It must handle operation histories like that shown in \Cref{fig:history_dag}.

\begin{figure}[t]
    \centering
    \includegraphics[width=0.95\columnwidth]{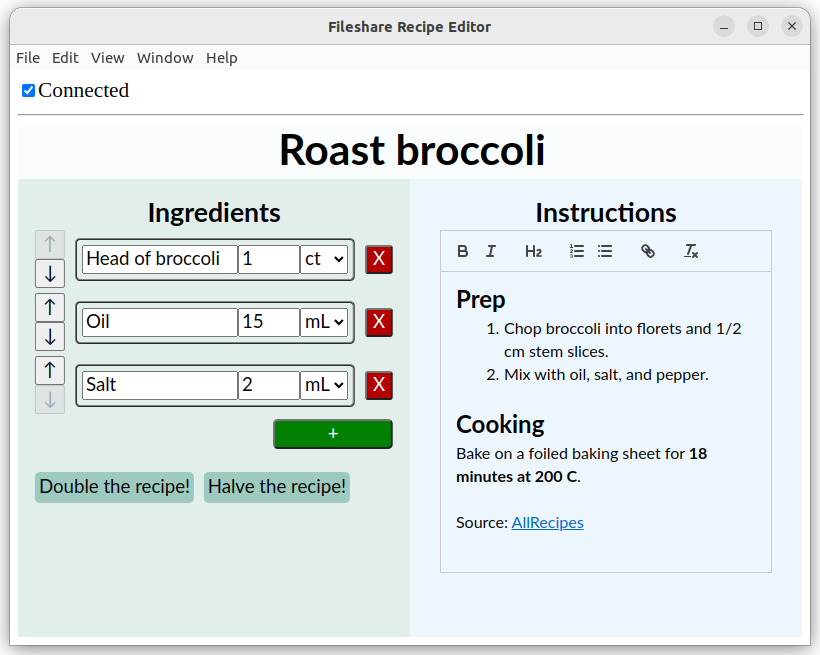}
    \caption{A collaborative recipe editor built on top of Collabs.}
    \label{fig:recipe_ui}
\end{figure}

\Cref{fig:recipe_ui} shows the app's user interface. Each GUI component stores its collaborative state in one of Collabs's built-in CRDTs: rich text for the instructions (\texttt{CRichText}), a list of objects for the ingredients list (\texttt{CList}), a last-writer-wins variable for each ingredient's units (\texttt{CVar}), etc.

\begin{figure}[t]
    \centering   \includegraphics[width=0.8\columnwidth]{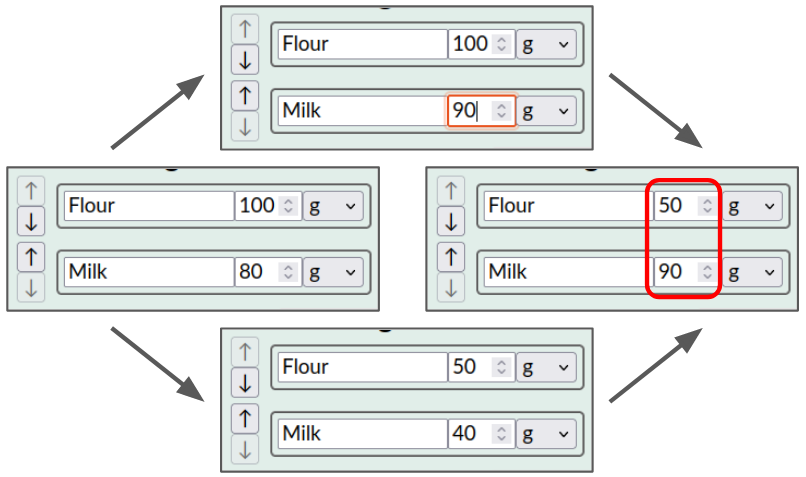}
    \caption{Potential anomalous behavior when one user adjusts a recipe while another halves it concurrently.}
    \label{fig:scale_anomaly}
\end{figure}

Our recipe editor also has a ``scale recipe'' button that scales all ingredients by a given amount---e.g., halving the recipe. A naive implementation of this operation would loop over each ingredient and set its amount to $(\textrm{current amount}) * (\textrm{scale factor})$. However, this suffers from an anomaly under concurrent edits, shown in \Cref{fig:scale_anomaly}: in a bread recipe, a user who adjusts the amount of milk to 90 grams could override a user who concurrently halves the recipe, leading to a bread recipe with far too much liquid (i.e., porridge).

The desired behavior is intuitively obvious---scale all ingredient amounts, including ones changed concurrently \cite{for_each}---but unusual enough that it is not built-in to Collabs. Instead, it needs a novel, app-specific CRDT algorithm that sends its own messages over the network.

\begin{figure}
    \centering
    \includegraphics[width=0.8\columnwidth]{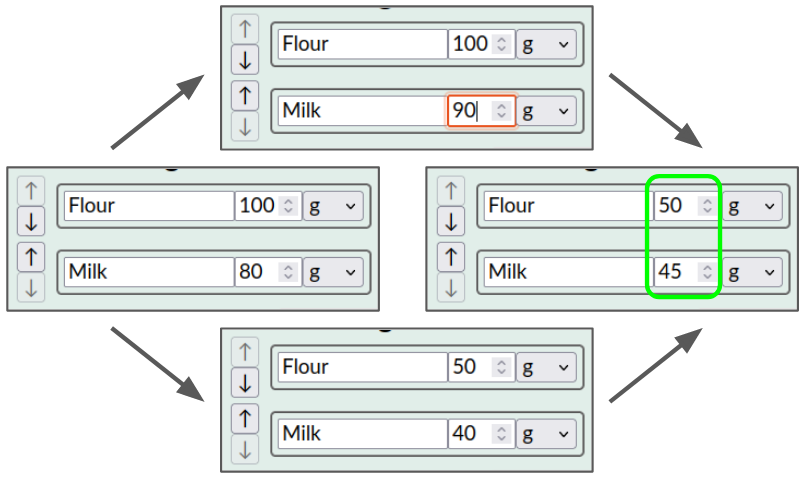}
    \caption{Actual behavior when one user adjusts a recipe while another halves it concurrently.}
    \label{fig:scale_correct}
\end{figure}

Collabs allows any developer to implement and use such custom CRDT algorithms. App developers can make and use app-specific CRDTs, while researchers can implement their own algorithms as third-party libraries. For our recipe editor, we define a class \texttt{CScaleNum} implementing a custom ``scalable number'' CRDT and use it to control each ingredient's amount. Thus our demo actually behaves like in \Cref{fig:scale_correct}: scaling the recipe also scales concurrent changes to the recipe.

\begin{figure}
    \centering
    \includegraphics[width=0.8\columnwidth]{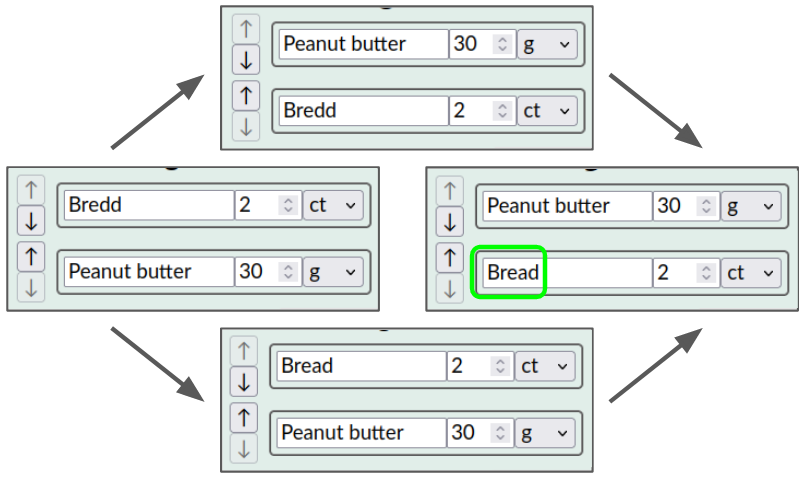}
    \caption{One user moves an ingredient while another edits it concurrently, fixing the typo ``Bredd''.}
    \label{fig:move_correct}
\end{figure}

Our recipe editor has fine-tuned semantics for other operations as well. First, if one user moves an ingredient while another user edits the ingredient concurrently, both changes are preserved, as shown in \Cref{fig:move_correct}.

Second, if one user deletes an ingredient while another user edits it concurrently, then the update ``wins'': the deletion is canceled, saving the second user's work.


Neither of these operations are possible with traditional list CRDTs. Instead, recent papers \cite{list_with_move, update_wins} describe how to implement them by ``composing'' existing CRDTs. Collabs lets you implement those compositional designs literally, then encapsulate them in reusable classes. We demonstrate this by implementing both papers' algorithms inside Collabs's built-in \texttt{CList} CRDT (see \Cref{app:composed_algs}). Because consistency properties also compose, the resulting \texttt{CList} is automatically strongly eventually consistent, and it has equivalent op- and state-based CRDT behaviors---two properties that are difficult to prove for traditional ``from-scratch'' CRDTs.

\vspace{-3mm}
\subsection{Semantic Flexibility as a Principle}
\label{sec:flexibility_principle}
\vspace{-2mm}

We believe that our framing of semantic flexibility as a desirable property is itself a contribution.

We just saw how semantic flexibility helps our collaborative recipe editor respect users' expectations and intents. Several other papers explore semantic choices for specific collaborative operations, including operations on rich text \cite{peritext, ignat_rich_text}, spreadsheet columns \cite{update_wins}, slideshow shapes \cite{for_each}, and shopping lists \cite{json_crdt}. Even early papers on CRDTs explore semantic choices like ``add-wins'' vs ``remove-wins'' sets \cite{state_based_patterns}. 

However, those papers are largely theoretical. Existing collaboration \emph{frameworks} have not adopted semantic flexibility as a design principle. Instead, they \emph{dictate} behaviors that apps can either take or leave; at best, these include some specific semantic choices from prior papers. 

In particular, a number of collaboration frameworks attempt to automatically derive CRDTs from sequential code (see \Cref{sec:related}). In our view, these automated techniques are undesirable because collaborative apps are fundamentally different from sequential apps. Just as it is not possible to derive a single-user app's business logic from its type signature, it is also not possible to derive the nuanced, concurrency-aware behavior we described above from sequential code. Instead, a programmer must think through concurrent scenarios and choose how their app should respond.

\vspace{-4mm}
\section{Collabs's Design}
\label{sec:system}
\vspace{-3mm}

To make a variety of collaborative apps and behaviors possible in Collabs---including rich-text editing, shared whiteboards, and the nuanced recipe editor seen above---we faced a number of challenges. These required a number of important design decisions, each of which enabled a different aspect of generality and reusability. The challenges include:
\vspace{-1mm}
\begin{description}
    \item[Flexibility, provided by the Collab API.] How can we support custom message-passing CRDTs like \texttt{CScaleNum}, with low performance overhead? 
    We propose \emph{the Collab API} to tackle this.
    \item[Composition, provided by the Tree of Collabs.] CRDT algorithms, like the \texttt{CText} text CRDT backing each ingredient's text field, are designed to work in isolation: they assume a single CRDT with full access to its own broadcast network and storage space. How can we make these algorithms work, unchanged, in a setting with many CRDTs, without compromising eventual consistency? How can we support ``composed'' CRDTs like \texttt{CList}, which are made up of many internal CRDTs? We use \emph{the tree of Collabs} to solve this.
    \item[Dynamism, provided by Parental Control.] Our recipe editor's data model is dynamic: the user can add and delete ingredients, changing the number and identities of the \texttt{CIngredient} CRDTs. How can we keep these dynamically-created CRDTs in sync across users? Furthermore, how can we support this dynamism in a generic way, allowing custom element types in collections (e.g., \texttt{CList<CIngredient>}) as well as novel, custom collection CRDTs in the future? We address this using \emph{parental control}.
\end{description}
\vspace{-2mm}

We now describe the core design techniques that Collabs uses to address the above challenges.

\subsection{Flexibility: The Collab API}
\label{sec:system1}
\vspace{-1mm}

Our first core technique is an API for a self-contained hybrid op-based/state-based CRDT, which we call a \emph{Collab} (short for ``collaborative data structure''). It is concretely implemented as the abstract class \texttt{Collab}. The Collab API solves the Flexibility challenge explained above.

\begin{figure}
    \centering
    \includegraphics[width=0.4\columnwidth]{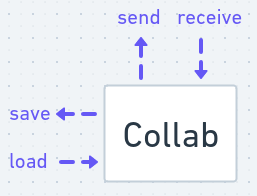}
    \caption{The Collab API.}
    \label{fig:collab_api}
\end{figure}

\Cref{fig:collab_api} illustrates the Collab API. It is a black box that sends and receives messages and that can save and load states.

Send and receive have the contract of an op-based CRDT: assuming that sent messages are broadcast to all of the Collab's replicas in causal order \cite{lamport_causal}, all replicas must converge to a consistent state. Save and load have the contract of a state-based CRDT: loading a saved state must be equivalent to receiving all of the updates that contributed to the saved state, ``merging'' them with existing updates.

For example, the op-based counter CRDT is a simple algorithm that sends ``+1'' messages and counts the total number of ``+1''s. A Collab implementing this algorithm does exactly that, as shown in \Cref{fig:counter_code}.\footnote{Collabs's actual \texttt{CCounter} Collab is more complex because it also supports state-based merging, which requires a more complex algorithm \cite{rdts}. It is still not much more complicated though, and it occupies $< 100$ lines of code.} The actual messages that Collabs sends over the network are only modestly larger than the literal ``+1''s (typically $\approx 50$ bytes), 
and likewise for CPU and memory usage. Thus we achieve performance flexibility: performance is dominated by the Collab implementation itself, which programmers can fully control.

\begin{figure}
\begin{lstlisting}[language=TypeScript]
class COpBasedCounter extends CPrimitive {
  private _value = 0;
  getValue() {
    return this._value;
  }
  increment() {
    super.sendPrimitive("+1");
  }
  protected receivePrimitive(message: Uint8Array | string, meta: MessageMeta) {
    assert(message == "+1");
    this._value++;
  }
  // Omitted: save and load of this._value.
}
\end{lstlisting}
\caption{A Collab implementing the op-based counter CRDT.}
\label{fig:counter_code}
\end{figure}

Note that the Collab contracts are not enforced programmatically, only described in our API docs. This is necessary to allow arbitrary CRDT implementations, including behaviors and optimizations that we have not anticipated.

\vspace{-4mm}
\subsection{Composition: The Tree of Collabs}
\label{sec:system2}
\vspace{-2mm}

Our second core technique is the \emph{tree of Collabs}. This is an implicit tree that is used to organize a document's Collab instances at runtime and that is replicated across all devices. It solves the Composition challenge described in \Cref{sec:system}.

\begin{figure}[t!]
    \centering
    \includegraphics[width=0.6\columnwidth]{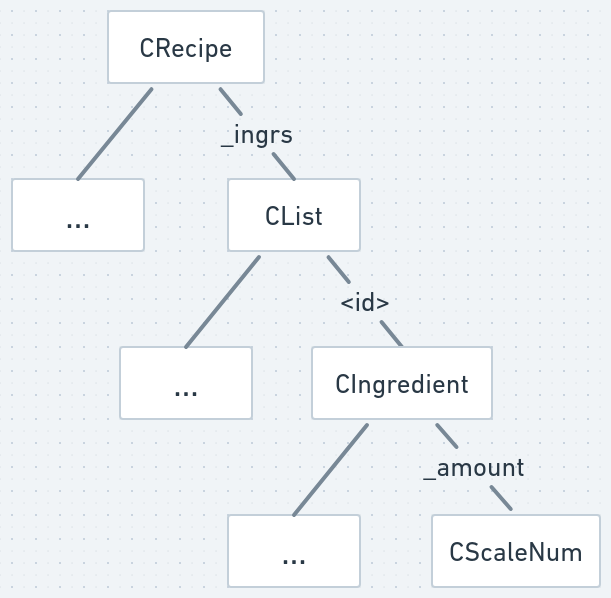}
    \caption{The tree of Collabs corresponding to a \texttt{CRecipe}.}
    \label{fig:tree_of_collabs}
\end{figure}

To explain this tree, \Cref{fig:tree_of_collabs} shows the tree corresponding to a recipe (class \texttt{CRecipe}) with three \texttt{CIngredients}. At the root is \texttt{CRuntime}, a built-in class that actually connects to a broadcast network and guarantees tagged reliable causal broadcast \cite{pure_op_based_crdts}. Below that, we have the app's top-level state, a \texttt{CRecipe}, which itself contains a \texttt{CList<CIngredient>} that stores the ingredients (among other children). Next, that \texttt{CList}'s children are its list elements, the \texttt{CIngredients}, which are dynamic in number.\footnote{We omit showing the internal structure of these Collabs, which sometimes involve further internal layers due to composition.}

The leaves of this tree are ``primitive'' Collabs that send and receive their own messages, usually in the form of Google Protobuf byte arrays. The internal nodes are more interesting: they can send their own messages, but they are also responsible for passing on their children's messages.

For example, when an ingredient's \texttt{\_amount} sends a message, the \texttt{CIngredient} sends that message onwards up the tree, tagged with a \emph{name} to multiplex among its children. These names are assigned in \texttt{CIngredient}'s constructor.

When \texttt{CIngredient} receives a message from a collaborator, it then demultiplexes the message and delivers it to the named child. This takes place in a reusable superclass called \texttt{CObject}; the \texttt{CIngredient} code in the actual app contains no CRDT logic. A similar pattern takes place up and down the tree, leading to the message flow in \Cref{fig:message_flow}.

\vspace{-4mm}
\subsection{Dynamism: Parental Control}
\label{sec:system3}
\vspace{-2mm}

Our third core technique is a design principle to support dynamic tree hierarchies that can evolve and change at runtime, like how the list of ingredients adds and deletes \texttt{CIngredient} CRDTs (its child nodes).

\begin{figure}[t]
    \centering
    \includegraphics[width=0.95\columnwidth]{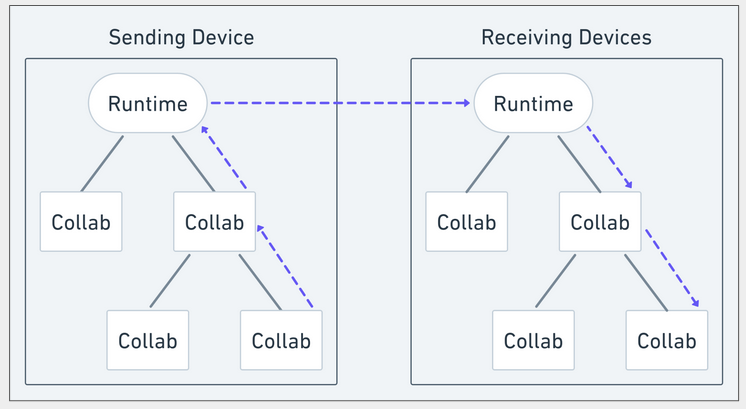}
    \caption{Message flow up and down the tree of Collabs.}
    \label{fig:message_flow}
\end{figure}

The design principle is \emph{parental control}: a parent node in the tree of Collabs is completely in charge of its own children. In particular, it can create and destroy children, and it can interfere with child messages (and saved states) as needed to ensure strong eventual consistency. This gives parents the flexibility to implement arbitrary collection semantics, lazy loading, complex composition techniques like the semidirect product of CRDTs \cite{semidirect}, etc.

To illustrate parental control, let us walk through the lifecycle of a \texttt{CIngredient} in the recipe's \texttt{CList<CIngredient>}. When the user adds an ingredient, the \texttt{CList} broadcasts a ``create'' message to its replicas. Upon receiving this message, each \texttt{CList} replica constructs a \texttt{CIngredient} using a callback registered in the list's own constructor (see \Cref{fig:construct_clist}). It then stores the \texttt{CIngredient} keyed by a unique new name included in the message.

\begin{figure}[t]
\begin{lstlisting}[language=TypeScript]
new CList(
  // Boilerplate used to set up the tree.
  ingrsInit,
  // Callback that constructs new list elements.
  (valueInit) => new CIngredient(valueInit)
);
\end{lstlisting}
\caption{Code to construct a \texttt{CList<CIngredient>}.}
\label{fig:construct_clist}
\end{figure}

Later, when the \texttt{CIngredient} sends a message, the \texttt{CList} sends that message onwards up the tree, tagged with the ingredient's name. Since the name is consistent across replicas, each receiving list knows to forward the message to its own replica of the ingredient, like in \Cref{fig:message_flow}.

Recall from \Cref{sec:flexibility} that our recipe editor has the option to use a delete-wins semantics instead of update-wins: deleted ingredients disappear permanently, even if they are edited concurrently. To implement a ``permanent'' deletion like this, the parent \texttt{CList} deletes its own replica of the \texttt{CIngredient} and sends a message telling other replicas to do likewise, reclaiming memory. If another user concurrently edits the ingredient, the \texttt{CList} chooses to ignore the resulting message: the delete ``wins'' over concurrent updates. This is allowed by parental control, since the list is free to interfere with its children's messages, including ignoring them outright.

Observe that each individual \texttt{CIngredient} is collaborative in the usual way: it believes that it is a single CRDT with access to its own broadcast network. Thus its state is strongly eventually consistent. The \texttt{CList}'s algorithm then ensures that the whole list is strongly eventually consistent, even in the presence of concurrent adds, deletes, and ingredient operations. In other words, consistency properties compose. They continue composing to arbitrary depths (lists of lists of …), allowing complex, dynamic tree hierarchies that are nonetheless consistent.

\vspace{-4mm}
\section{Complex CRDTs through Composition}
\label{sec:composition}
\vspace{-3mm}

Recall from \Cref{fig:recipe_ui} that a recipe in our recipe editor app has a title, ingredients, and instructions. In a non-collaborative version of the app, you might model a recipe using a class \texttt{Recipe} with instance variables \texttt{\_title}, \texttt{\_ingrs}, and \texttt{\_instrs}. The Collabs version of this data model, with fully collaborative operations, is quite similar:

\begin{lstlisting}[language=TypeScript]
class CRecipe extends CObject {
  private _title: CVar<string>;
  private _ingrs: CList<CIngredient>;
  private _instrs: CRichText;

  addIngredient(index: number) {
    this._ingrs.insert(index);
  }
  ...
}
\end{lstlisting}

Here \texttt{CRecipe} is a custom ``CRDT object'' defined through composition: it wraps a conjunction of other CRDTs in its own API. \texttt{CObject} is a superclass for these CRDT objects. Observe that \texttt{CRecipe.\_ingrs} is itself composed: it is our built-in \texttt{CList} applied to elements of type \texttt{CIngredient}, which is another custom CRDT object.

\texttt{CObject} and \texttt{CList} guarantee that correctness properties compose. Then \texttt{CRecipe} automatically satisfies the two guarantees from \Cref{sec:model}---strong convergence, and equivalent op- and state-based semantics---because its components do.

Composed and encapsulated data models like \texttt{CRecipe} let a Collabs programmer build up useful but nontrivial CRDT semantics, with a simple external API. We give further example code in \Cref{app:composition}, including a list-with-move CRDT \cite{list_with_move} implementing the intuitive move semantics in \Cref{fig:move_correct}.

\vspace{-4mm}
\section{Capabilities Evaluation}
\label{sec:eval_cap}
\vspace{-3mm}

As a collaboration framework, Collabs is designed to let programmers implement their own collaborative apps. We are interested in evaluating how well it succeeds at this goal, i.e., the quality of its programming model. This is difficult to evaluate objectively because concrete metric like lines-of-code are not always reliable \cite{loc_unreliable}.

Instead, we compare specific capabilities of Collabs with selected collaboration frameworks. These capabilities determine the kinds of apps that are possible or easy to program on top of a given framework. Additional frameworks are described in our Related Work (\Cref{sec:related}). 

\Cref{tab:capabilities} summarizes the capabilities of the compared frameworks. A \yes{} indicates full support for a given capability; \no{} indicates no support; and \half{} indicates limited support (see \Cref{app:capabilities} for justifications). Briefly, the capabilities are:
\vspace{-2mm}
\begin{description}
    \item[Local-first] The framework is local-first in the sense of Kleppmann et~al.~\cite{local_switch}.
    \item[Rich-text editing] The framework has built-in support for collaborative rich-text editing.
    \item[Nested data] The framework supports arbitrarily nested data.
    \item[List-with-move] The framework supports moving elements in lists with the intuitive semantics shown in \Cref{fig:move_correct}.
    \item[Encapsulated data models] The framework lets an app define encapsulated, type-safe data models for portions of its own state.
    \item[Semantic flexibility \& performance flexibility] The framework permits both semantic flexibility and performance flexibility as defined in the Introduction.
\end{description}
\vspace{-2mm}

\begin{table*}[t]
    \centering
    \renewcommand{\arraystretch}{1.1}
    \addtolength{\tabcolsep}{-0.2em}
    \footnotesize
    \begin{tabular}{m{3cm}|ccccccc}
         & Collabs & Yjs \cite{yjs} & Automerge \cite{automerge} & Legion \cite{legion} & ShareDB \cite{sharedb} & OWebSync \cite{owebsync} & Replicache \cite{replicache} \\ \hline
         Local-first & \yes{} & \yes{} & \yes{} & \yes{} & \half{} & \half{} & \half{} \\
         Rich-text editing & \yes{} & \yes{} & \yes{} & \no{} & \yes{} & \no{} & \no{} \\
         Nested data & \yes{} & \yes{} & \yes{} & \no{} & \yes{} & \yes{} & \no \\
         List-with-move & \yes{} & \half{} & \half{} & \no{} & \no{} & \no{} & \no{} \\
         Encapsulated data models & \yes{} & \half{} & \no{} & N/A & \no{} & \no{} & N/A \\
         Semantic flexibility \& performance flexibility & \yes{} & \no{} & \no{} & \no{} & \half{} & \no{} & \yes{}
    \end{tabular}
    \normalsize
    \caption{Collabs's capabilities compared to other collaboration frameworks.}
    \label{tab:capabilities}
\vspace{-6mm}
\end{table*}

A separate preliminary paper describes real programmers' experience using Collabs, Yjs, and Automerge to add collaboration to simple apps \cite{yicheng_user_study}. 

\vspace{-4mm}
\section{Performance Evaluation}
\label{sec:eval_perf}
\vspace{-3mm}

\subsection{Many-User Rich-Text Editing}
\label{sec:rich_text_perf}
\vspace{-2mm}

We evaluated Collabs's performance using a collaborative rich-text editing benchmark.\footnote{Source code is available at \url{https://github.com/composablesys/collabs-rich-text-benchmarks}. We also provide a \href{https://cmu.box.com/s/ina8bc35c2wyu22q0tydfyaeb2c4744c}{full data download}.
} This is a popular but difficult collaborative app. It is especially difficult to scale to a large number of users. For example, Google Docs limits a document to 100 simultaneous editors \cite{google_docs_100}, and experiments with the 2016 version revealed high end-to-end latencies well before the then-limit of 50 editors \cite{dang_ignat_rtc_perf}.

We implemented a collaborative rich-text editor using Collabs's built-in CRDTs, then asked: how many simultaneous users can this editor support before the user experience breaks down, for realistic workloads? How does that compare to similar editors built on top of other collaboration frameworks, and what are the bottlenecks to further scaling?

The collaborative rich-text editor is a TypeScript webpage using the Quill rich-text editor GUI v1.3.7
with a basic set of allowed formats (bold, italic, block headers 1 \& 2, and lists). \Cref{fig:quill_screenshot} in the appendix shows a screenshot.

Our benchmark's structure and code are based on OWebSync's eDesigners benchmark \cite{owebsync}.

\vspace{-5mm}
\subsubsection{Experimental Setup}
\vspace{-2mm}

Our benchmarks used a central server to connect collaborators. (Although not necessary for Collabs, this setup is simple and allows us to compare to server-based frameworks.) Each benchmark featured 16--144 users editing a shared document simultaneously. The server was an AWS EC2 t2.medium instance running Ubuntu 22.04.3 and Node.js v18.17.1. Each user was a Docker container running on an AWS Fargate spot instance. For geo-distribution, the server was deployed in AWS's us-east-1 region, while users were evenly divided between regions us-west-1 (60 ms from us-east-1) and eu-north-1 (110 ms from us-east-1) \cite{aws_latencies}.

Each user opened the server's webpage using Headless Google Chrome v116.0.5845.96. We used Puppeteer
to control each user's keyboard and mouse inputs, simulating a realistic workload. Specifically, each user typed at 6 keystrokes per second, with the content, cursor movements, and operations (insert/delete) drawn from a recorded typing trace \cite{automerge_perf}. Also, $1\%$ of operations were rich-text formatting operations: either using Ctrl+b/i to change the format under the cursor ($0.5\%$), changing the format of a random 1-50 character range of text near the cursor ($0.25\%$), or clicking a toolbar button to change the current paragraph's block type ($0.25\%$).

We ran each experiment three times and report averages.

\vspace{-4mm}
\subsubsection{Collaboration Frameworks}
\vspace{-2mm}

The collaboration frameworks we evaluated are:
\vspace{-2mm}
\begin{description}
    \item[Collabs] Collabs v0.13.4's built-in \texttt{CRichText} CRDT, which implements a variation of Peritext \cite{peritext} on top of the Fugue list CRDT \cite{fugue}. For networking, we used Collabs's @collabs/ws-client and @collabs/ws-server plugins, which provide a basic WebSocket client/server.
    \item[CollabsNoVC] Same, but with the vector clock entries from \Cref{sec:model} disabled using an option in Collabs's runtime. We describe the rationale for this later, in \Cref{sec:analysis}.
    \item[Yjs \cite{yjs}] The Yjs library v13.6.7's \texttt{Y.Text} CRDT, Yjs's y-quill Quill integration, and its y-websocket WebSocket client/server. 
    Of note, Yjs has a reputation amongst developers for fast collaborative text editing
    \cite{yjs_perf_blog_post, dmonad_benchmarks}.
    \item[Automerge \cite{automerge}] The Automerge library v2.1.2-alpha.0's string CRDT, a Quill integration we implemented ourselves, and a simple WebSocket echo server.\footnote{At the time of writing, Automerge did not have a rich-text editor integration, and the provided automerge-repo-sync-server was less scalable.
    }
    \item[ShareDB \cite{sharedb}] The ShareDB framework v4.0.0 and its rich-text type v4.1.0. ShareDB uses Operational Transformation instead of CRDTs, similar to Google Docs. We followed an official example to setup its Quill integration and WebSocket client/server.
    \item[GDocs] A Google Doc shared via an editable link. This is not a perfect comparison because Google Docs has many more features than Quill and a different server environment, but it is still an interesting comparison. Internally, it uses the Jupiter OT algorithm \cite{jupiter}. 
\end{description}

\vspace{-6mm}
\subsubsection{Experiments}
\vspace{-2mm}

In initial experiments, we saw that the Quill rich-text editor was a client-side CPU bottleneck for modest numbers of users. To determine the limits of the collaboration frameworks, independent of this bottleneck, we ran two sets of experiments.

For the first set of experiments, \textbf{OptQuill}, we used Quill with the following optimizations:
\vspace{-2mm}
\begin{enumerate}
    \item Instead of updating Quill's state every time it receives an update from the server, each user's client applies updates in a batch at most once per 50 ms. This reduces the amount of time spent in render-related code without noticeable downsides.
    \item We disabled shared cursor indicators, which are expensive to update in the display.
    \item For CRDT frameworks with $\ge 64$ users, instead of sending local updates to the server immediately, each client sends updates in batches at most once per second. (ShareDB does a similar optimization by default---see \Cref{sec:analysis}.) Empirically, this causes other clients to spend less time in Quill code, but it also confounds our measurements by increasing end-to-end latency and reducing network usage. \label{item:opt3}
\end{enumerate}
\vspace{-2mm}
Each user's container was allocated 1 AWS vCPU and 2 GiB of memory, representing a modest device.

For the second set of experiments, \textbf{NoQuill}, we used a lightweight simulation of Quill that does not actually render its state. This allowed us to omit OptQuill's confounding optimization \ref{item:opt3}. Each user's container was allocated 0.5 vCPU and 1 GiB of memory: nominally, the other 0.5vCPU and 1 GiB of memory were reserved for an imaginary GUI.

\vspace{-3mm}
\subsubsection{Results: OptQuill}
\vspace{-2mm}

We ran each experiment for six minutes and report statistics for the final minute. The first five minutes serve to create a document with several pages of formatted text and a rich edit history. Our main question is whether the collaborative editor is still functional during the sixth minute: (1) are users' clients responsive to their own operations, and (2) do those operations show up for other users?

\begin{figure}[t!]
    \centering
    \includegraphics[width=\columnwidth]{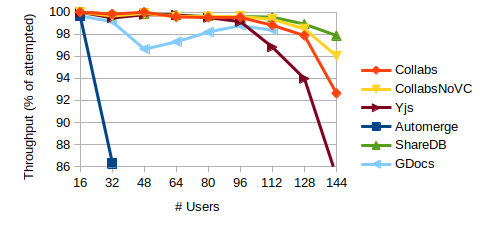}
    \vspace{-8mm}
    \caption{Throughput as a function of the number of users. Google Docs' increasing trend for $\ge 48$ users likely reflects the lower rate of remote updates (the server was overwhelmed).}
    \label{fig:throughput}
    \vspace{1mm}
\end{figure}

\begin{figure*}[t]
    \centering
    \begin{subfigure}[b]{0.44\textwidth}
        \centering
        \includegraphics[height=1.6in]{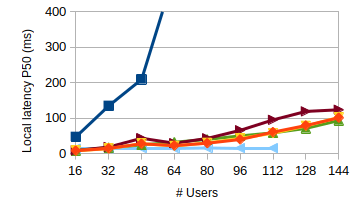}
        \vspace{-3mm}
        \caption{Median}
        \label{fig:local_latency_p50}
    \end{subfigure}
    \begin{subfigure}[b]{0.55\textwidth}
        \centering
        \includegraphics[height=1.6in]{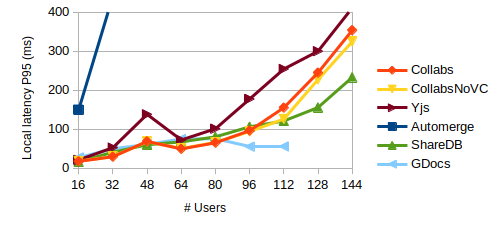}
        \vspace{-3mm}
        \caption{95th percentile.}
        \label{fig:local_latency_p95}
    \end{subfigure}
    \caption{Local latency as a function of the number of users.}
    \label{fig:local_latency}
\vspace{-3mm}
\end{figure*}

\begin{figure*}[t]
    \centering
    \begin{subfigure}[b]{0.44\textwidth}
        \centering
        \includegraphics[height=1.6in]{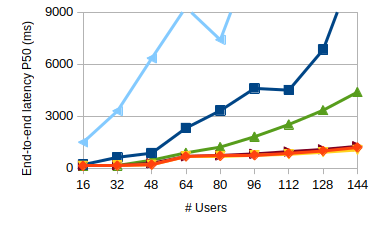}
        \vspace{-3mm}
        \caption{Median}
        \label{fig:e2e_latency_p50}
    \end{subfigure}
    \begin{subfigure}[b]{0.55\textwidth}
        \centering
        \includegraphics[height=1.6in]{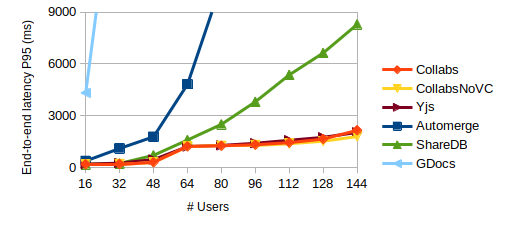}
        \vspace{-3mm}
        \caption{95th percentile.}
        \label{fig:e2e_latency_p95}
    \end{subfigure}
    \caption{End-to-end latency as a function of the number of users. The underlying network latency has median 170 ms and 95th percentile 220 ms; for $\ge 64$ users, the least possible end-to-end latency is higher due to sender-side batching (optimization \ref{item:opt3}).}
    \label{fig:e2e_latency}
    \vspace{-4mm}
\end{figure*}

\begin{table*}[t]
    \centering
    \renewcommand{\arraystretch}{1.1}
    \addtolength{\tabcolsep}{-0.2em}
    \small
    \begin{tabular}{lcccccc}
        Framework & \shortstack{E2E latency, \\ median (ms)} & \shortstack{E2E latency, 95th \\ percentile (ms)} & \shortstack{Server \\ CPU (\%)} & \shortstack{Client \\ CPU (\%)} & \shortstack{Client \\ memory (MiB)} & \shortstack{Client \\ network (KiB/sec)} \\ \hline
        Collabs & 864 & 1436 & 10 & 80 & 345 & 187 \\
        CollabsNoVC & 810 & 1373 & 9 & 79 & 345 & 39 \\
        ShareDB & 2529 & 5367 & 101 & 81 & 333 & 36
    \end{tabular}
    \normalsize
    \caption{Performance comparison for Collabs and ShareDB at 112 users (OptQuill experiments). Resource usages are averages for the final minute. Client CPU and memory are for the Chrome process; client network is send+receive for the client's container.}
    \label{tab:sharedb_vs_collabs}
\end{table*}

Figures \ref{fig:throughput}--\ref{fig:e2e_latency} answer these questions as a function of the number of users, using data from the \textbf{OptQuill} experiments (and Google Docs). \Cref{fig:throughput} shows \emph{throughput}: the number of operations that users were able to perform, as a percentage of those attempted. This declines when the client is overwhelmed, since our Puppeteer script waits to type/click the next operation until after the browser acknowledges the previous input.

\Cref{fig:local_latency} shows \emph{local} latency: how long it takes for keystrokes to render on the user's own screen. Although all frameworks use local-first or optimistic updates to render a user's own keystrokes without a round-trip to the server, there can still be client-side delays: in particular, if the collaboration framework (or Quill) blocks the main thread with long synchronous tasks. The first line at 100ms is a folklore threshold for responsive GUI interactions.

Finally, \Cref{fig:e2e_latency} shows \emph{end-to-end latency}: the time from when one user performed an operation until it showed up for other users. This typically increases when the server is busy. To measure end-to-end latency, once every 10 seconds, each user typed a \emph{sigil}: a unique sequence of 5 characters. Each user logs when they initiate a sigil's final keypress and when\footnote{All devices' clocks were synchronized using the Amazon Time Sync Service. Clock error bounds all measured $< 1.5$ ms from UTC.} they see any sigil render;\footnote{Once the sigil has been added to the page's HTML, we schedule a \texttt{window.requestAnimationFrame} callback, which schedules a new task using \texttt{setTimeout}; we log the sigil as rendered during that task. This is guaranteed to be after the next render, but may be delayed \cite{w3c_event_loop}.

For Google Docs, we instead use \texttt{MutationObserver} to monitor an HTML copy of the text that is provided to certain Chrome extensions.} end-to-end latency is the difference. The figure shows both median and 95th percentile latencies, aggregated across the entire minute.


\begin{table}[t]
    \centering
    \renewcommand{\arraystretch}{1.1}
    \addtolength{\tabcolsep}{-0.2em}
    \small
    \begin{tabular}{lcc}
        Framework & \shortstack{Max users \\ (OptQuill)} & \shortstack{Max users \\ (NoQuill)} \\ \hline
        Collabs & 112 & 80 \\
        CollabsNoVC & 112 & 112 \\
        Yjs & 96 & 48 \\
        Automerge & 16 & - \\
        ShareDB & 112 & 112 \\
        GDocs & 16 & N/A
    \end{tabular}
    \normalsize
    \caption{The maximum number of users for which each framework was fully functional.}
    \label{tab:max_users}
\end{table}

Let us conservatively consider a framework to be ``fully functional'' if it achieves:
\vspace{-2mm}
\begin{itemize}
    \item $\ge 98$\% throughput;
    \item $\le 100$ ms median local latency;
    \item $\le 3$ second median end-to-end latency; and
    \item 95th percentile latencies at most twice as large.
\end{itemize}
\vspace{-2mm}
With this definition, we have the maximum fully functional user counts shown in the OptQuill column of \Cref{tab:max_users}.

For larger user counts, the CRDT libraries see decreasing throughput, while ShareDB experiences increasing remote latencies. This matches the stereotype that CRDTs are client CPU-intensive while Operational Transformation is server CPU-intensive \cite{ot_complexity}. Our CPU usage statistics bear this out.

Although Collabs and ShareDB are both functional at 112 users, they have different resource usages; see \Cref{tab:sharedb_vs_collabs}. In particular, the Collabs variants have lower end-to-end latency and server CPU usage than ShareDB, while other resource usages are comparable (except Collabs's network usage, discussed in \Cref{sec:analysis}). This suggests a better user experience and lower deployment costs for Collabs.

\vspace{-4mm}
\subsubsection{Results: NoQuill}
\vspace{-2mm}

We also investigated the maximum ``fully functional'' user counts for \textbf{NoQuill}. The results are broadly similar to OptQuill except for the following differences (see \Cref{tab:max_users}).

For Collabs, at 96 users, remote updates were not delivered reliably. This is because, without sender-side batching (OptQuill's optimization \ref{item:opt3}), clients overwhelm the server's network link. 
CollabsNoVC avoids this problem for reasons discussed in \Cref{sec:analysis} below.

Automerge achieves only 64\% throughput even at 16 users. Relative to OptQuill, replacing Quill's CPU usage with a smaller CPU allocation (0.5 vCPU) harms performance: OptQuill spends more time in Automerge code than Quill code.\footnote{Indeed, for OptQuill with 16 users, Automerge's average client CPU usage is over twice that of ShareDB, whose CPU usage is mostly Quill. Mysteriously, our recorded CPU profiles appear to disagree; this may reflect differences between clients, since we only recorded one client per trial. 
}

For Yjs with $\ge 64$ users, end-to-end latency varies widely throughout trials, often exceeding 10 seconds. Periods of high latency correspond to high server CPU usage ($> 50$\%); we believe that, without sender-side batching, the server becomes overloaded by many small messages. Note that unlike Collabs's and Automerge's servers, Yjs's server chooses to act as a CRDT replica, increasing its CPU usage.

\vspace{-4mm}
\subsubsection{Analysis}
\label{sec:analysis}
\vspace{-2mm}

For each collaboration framework, we analyzed why it became non-functional at high user counts. Our conclusions are based on generic performance metrics (CPU, memory, network), Chrome CPU profiles recorded at the end of each experiment, and time-series data showing how each metric varied throughout the six minutes.

\paragraph{Collabs}
We already mentioned that in NoQuill with $\ge 96$ users, Collabs overwhelms the server's (virtualized) network link, which maxes out at $\approx 350$ Mib/sec. Indeed, the server's network traffic scales as $O(n^3)$ ($n$ = number of users) because average message size scales as $O(n)$. This is caused by the vector clock entries that Collabs uses to enforce causal-order delivery (\Cref{sec:model}): with $n$ simultaneous active users and sufficient network latency, there are sets of $\Omega(n)$ pairwise-concurrent messages, which mathematically require $\Omega(n)$-size logical timestamps for causality tracking \cite{vc_lower_bound}. This lower bound holds even though Collabs only sends causally-maximal vector clock entries, i.e., we avoid sending entries for inactive replicas.

Actually, these vector clock entries are only necessary in certain decentralized scenarios. In the experiments' setup, Collabs's WebSocket client/server already guarantee causal-order delivery. Thus it is safe to disable vector clock entries, yielding CollabsNoVC.


In OptQuill, sender-side operation batching (optimization \ref{item:opt3}) reduces the impact of vector clocks: messages in the same batch share vector clock keys. This allows scaling up to 112 users. At 128 users, clients become CPU bound, but more by Quill than Collabs: Chrome CPU profiles show a 1:2 ratio for Collabs:Quill computation time. 

An alternate solution would use a different network topology to broadcast messages, so that no single node bottlenecks.

\paragraph{CollabsNoVC}
We described the rationale for disabling vector clocks above. For OptQuill at 128 users, CollabsNoVC clients are again CPU bound: long synchronous tasks increase 95th percentile local latency.

For NoQuill, one might expect 128 users to have fully functional clients, but instead the server becomes CPU bound. We attribute this to the large number of small messages that it must broadcast, due to no sender-side operation batching.

\paragraph{Yjs}
Yjs became client CPU-bound at 112 users. We observed that the time spent processing remote updates scaled not with the number/rate of updates, but instead with the document's size: more users create a larger and more complex document during the first five minutes, reducing throughput in the final minute.

Chrome CPU profiles show that when processing a batch of received updates, the majority of Yjs's time is spent in its functions \texttt{YTextEvent.delta()} and \texttt{cleanupYTextFormatting()}. The former computes the changes from a batch of received updates, so that Yjs can update Quill; the latter cleans up the internal representation of formatting spans after a batch of updates. 
Both functions walk Yjs's entire linked-list representation of the rich-text state, which becomes large after many formatting operations.

We validated this explanation with a microbenchmark: a single user inserts 16.5k characters into a \texttt{Y.Text} CRDT from left to right, with alternating bold/non-bold formatting. Yjs's linked list representation then has 32k elements, like the document at the end of OptQuill's 112 user benchmark (33k elements). Afterwards, calling \texttt{YTextEvent.delta()} for a future update takes $\approx 5$ ms. Thus calling that method twice and performing other work (as y-quill does) consumes a sizeable fraction of the 50 ms between remote batches.

\paragraph{Automerge}
Similar to Yjs, Automerge becomes CPU bound once the document grows large enough, due to time spent processing remote updates. Chrome CPU profiles show that the majority of Automerge's time is spent in code concerning formatting spans (marks).

In microbenchmarks, we found that the time required to process a single remote update scales with the document's ``complexity'', but not directly with the number of past operations. For example, the saved state from OptQuill's 32 user experiment is approximately twice as complex as that from the 16 user experiment, and applying a single remote update to that state takes several times as long: e.g., $2.0$ vs $0.5$ ms for a single-character insertion. However, if we use a single user to create a document comparable to the 32 user state---with a similar number of internal ops (54k) and formatting spans (900)---then applying a single-character insertion takes only $0.5$ ms. We could not determine the source of this difference.

\begin{figure*}
    \centering
    \begin{subfigure}[b]{0.44\textwidth}
        \centering
        \includegraphics[height=1.6in]{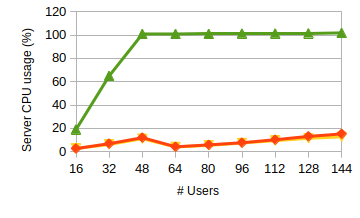}
        \vspace{-4mm}
        \caption{Server process CPU usage.}
        \label{fig:sharedb_server_cpu}
    \end{subfigure}
    \begin{subfigure}[b]{0.55\textwidth}
        \centering
        \includegraphics[height=1.6in]{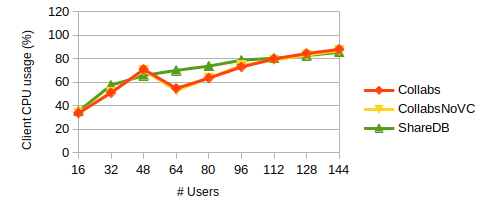}
        \vspace{-4mm}
        \caption{Client process CPU usage (average).}
        \label{fig:sharedb_client_cpu}
    \end{subfigure}
    \caption{ShareDB and Collabs CPU usages. Note that each server is a Node.js program, hence mostly single-threaded. Collabs's decrease at 64 users is likely due to sender-side batching (OptQuill's optimization \ref{item:opt3}).}
    \label{fig:sharedb_cpus}
    \vspace{-4mm}
\end{figure*}

\paragraph{ShareDB}
Unlike the CRDT libraries, ShareDB was primarily limited by server CPU usage. Operational Transformation's algorithmic complexity per op generally scales with the amount of concurrency \cite{ot_complexity}, so we should expect server CPU usage to scale as $O(n^2)$ ($n$ is the number of concurrent users). \Cref{fig:sharedb_server_cpu} is compatible with this, except that server CPU usage levels off at 100\% for 48 users.


Surprisingly, ShareDB continues to be functional well beyond 48 users, with only modestly increasing end-to-end latencies (\Cref{fig:e2e_latency}). To accomplish this, ShareDB performs sender-side operation batching---the inspiration for OptQuill's optimization \ref{item:opt3}. Specifically, a ShareDB client will only have one message ``in flight'' (waiting on the server) at a time \cite{seph_email}. Thus at high server loads, clients send updates in batches, with an adaptive batch size.

Because of how users type, each batch is usually a contiguous text insertion, which the ShareDB server handles as easily as a single keystroke. However, batching comes at the cost of higher end-to-end latency, since clients delay sending operations to the server. This effect is clearly seen in \Cref{fig:e2e_latency}: ShareDB's end-to-end latency increases with a steeper slope than the CRDT libraries'.

We might expect Collabs's lower server CPU usage to trade off against higher client CPU usage \cite{ot_complexity}. However, \Cref{fig:sharedb_client_cpu} shows that this effect is too small to be noticeable. Indeed, most CPU usage is devoted to Quill, and ShareDB clients must also do some work to transform local operations. Even in NoQuill 112 users---where Quill is not present and only ShareDB uses sender-side batching---ShareDB clients average 26\% CPU usage while CollabsNoVC averages 31\%.

\paragraph{GDocs}
We have less insight into Google Docs because we did not control the server. However, it appears to be limited by the server, like ShareDB and as expected for Operational Transformation. End-to-end latency exceeds our thresholds even for 32 users (median 3.3 seconds, 95th percentile 19.4 seconds). These results are similar to Dang and Ignat's \cite{dang_ignat_rtc_perf}.

To validate these latencies, we interacted with the Google Doc ourselves during a repeat of the 32 user experiment. Edits to one window took 5+ seconds to show up in another window on the same device, and edits from the experiment clients showed up in multi-second batches.

At larger user counts, attempting to view the document often resulted in "Trying to connect" or "Reconnecting" messages. We believe this affected the experiment clients as well: across all experiment trials, the maximum number of users who managed to actually type in the document never exceeded 51 (see \Cref{fig:gdocs_max} in the appendix).

We caution that our experiments are not a perfect comparison because Google Docs has many more features than our Quill editor (including shared cursors). Also, we hypothesize that a single Google Doc is allocated less server resources than our experiment server. That could explain why ShareDB scaled much better despite its similar architecture.

\vspace{-4mm}
\subsection{Performance: Metadata Overhead}
\label{sec:metadata_perf}
\vspace{-2mm}
Our rich-text editing benchmark in \Cref{sec:rich_text_perf} mostly evaluated the \emph{speed} of Collabs's op-based CRDT operations. However, a common criticism of CRDTs is that they store too much metadata, increasing memory usage and storage.

To evaluate this metadata overhead, we performed microbenchmarks where we load and save the saved states from the end of each experiment. We then recorded the saved state size, time to load and save the state, and resulting change in memory usage. For full info, see \Cref{app:metadata_perf}.

Briefly, we found that for the saved state at the end of the 112 user benchmark, Collabs has an acceptable metadata overhead. Its saved state is 557 KiB, only $2.3 \times$ as large as Quill's non-collaborative representation of the state; it loads in 330 ms and saves in 22 ms; and it occupies only $10.5$ MiB of memory. These are all practical for browser apps.

\vspace{-4mm}
\subsection{Discussion}
\vspace{-2mm}
We chose to benchmark collaborative rich-text editing because it is a prime use case that is also strenuous. Additionally, we focused on the extreme case of many-user all-active collaboration because it is a known challenge \cite{dang_ignat_rtc_perf}, and to ensure that our results are a ``lower bound'' on real-world scalability.

However, we stress that Collabs is not designed solely for rich-text or for large, active groups. Our other demos include the recipe editor described in \Cref{sec:flexibility}, a shared whiteboard, a cooperate minesweeper game, and a custom tensor-average CRDT for federated learning; and Collabs's state-based CRDT usage is suited for peer-to-peer collaboration and cross-device sync. Microbenchmark data for non-rich-text CRDTs (maps, counters, nested todo-lists, etc.) can be found in our full data download.

Nonetheless, it is encouraging to see that Collabs scales to over 100 rich-text editor users. This exceeds Google Docs' documented 100 user limit \cite{google_docs_100}, which we saw was optimistic when all users are active (\Cref{fig:gdocs_max}).

We identify three reasons for Collabs's good rich-text editing performance, based on our analysis in \Cref{sec:analysis}:

\textbf{Reason 1.} The Collab API permits more algorithms, which lets one choose better algorithms. In particular, a custom Collab can lay out its internal state using arbitrary data structures. Thus one does not need to store each CRDT's state as a linked list of items like in Yjs, which we saw makes it expensive to compute changes from received updates; and one does not need to store each character separately like in Automerge, which leads to large metadata overhead (\Cref{app:metadata_perf}).

\textbf{Reason 2.} Our built-in list and rich-text CRDTs are performant. This is mostly a result of their specific implementations, but it is related to Collabs's flexible and modular architecture: that let us re-implement our list CRDT several times without tearing out the rest of the library. 

\textbf{Reason 3.} Our choice of CRDTs instead of centralized Operational Transformation let us avoid server bottlenecks. This possibility was already raised in 2011 by Ahmed-Nacer et~al.\ \cite{crdt_perf_2011}, and when Dang and Ignet identified scalability bottlenecks with 50-user Google Docs in 2016, they cited CRDTs as a possible workaround \cite{dang_ignat_rtc_perf}. Our experiments show that Collabs lives up to this promise: in our setup, it can support over 100 simultaneous active users; its metadata overhead is acceptable; and it has noticeably lower CPU usage and end-to-end latency than ShareDB and Google Docs.



\vspace{-3mm}
\section{Related Work}
\label{sec:related}
\vspace{-1mm}

\paragraph{Local-First CRDT Libraries}
The most similar prior works to Collabs are Yjs and Automerge, two web CRDT libraries. These are the only collaboration frameworks we are aware of with a demonstrated ability to support real local-first apps. Like Collabs, both libraries support collaboration on rich text and arbitrarily nested data, provide networking and storage plugins alongside their network-agnostic CRDTs, and are actively-maintained open-source projects.

Unlike Collabs, neither library supports semantic flexibility or performance flexibility. Instead, they provide menus of built-in CRDTs. Automerge provides a single JSON CRDT that also supports counter and rich-text fields. Yjs provides composable list, map, text, and XML CRDTs. Also, neither library fully supports encapsulated data models like our \texttt{CRecipe} (\Cref{sec:composition}), although Yjs's author is interested \cite{yjs_jupyter_blog_post}.

Additionally, our collaborative rich-text editing benchmark shows that Collabs has better scalability than Yjs and Automerge on this popular application.

\paragraph{Additional Collaboration Frameworks}
Several academic works describe CRDT libraries for collaborative apps. Legion provides basic CRDTs but does not support nested data \cite{legion}. OWebSync provides a JSON CRDT with a novel state-based sync algorithm \cite{owebsync}. Neither library supports rich text, semantic flexibility, or performance flexibility.

Flec \cite{flec_nested} and ReScala's ARDTs \cite{rescala_ardts} were developed concurrently to Collabs and share some design decisions. Flec uses a similar tree of CRDTs; ARDTs support map and list composition similar to our \texttt{CObject} and \texttt{CList}. Both permit programmer-defined CRDTs, but with more restrictions. 


Other collaboration frameworks use server-based approaches instead of CRDTs. These are designed for traditional live collaboration; unlike Collabs, they do not support peer-to-peer collaboration. ShareDB uses Operational Transformation (OT) and supports JSON and rich text \cite{sharedb}. Replicache \cite{replicache} and Fluid Framework \cite{fluid} use server reconciliation, in which a central server assigns a total order to operations. Replicache uses programmer-defined operations, while Fluid Framework supports basic collaborative data structures. 

MRDTs \cite{mrdts}, ECROs \cite{ecros}, Katara \cite{katara}, and OpSets \cite{opsets} attempt to automatically derive CRDTs from sequential code. As we argued in \Cref{sec:flexibility_principle}, these automated techniques conflict with semantic flexibility.
Existing automated frameworks also lack practical features like rich-text CRDTs or the ability to optimize created CRDT implementations (perf  flexibility).

MRDTs interpret an abstract data type in terms of sets and replaces those sets with a specific set CRDT. OpSets apply sequential operations in Lamport-timestamp order. These output a single CRDT semantics per sequential data structure; it is unclear how to achieve concurrency-aware behavior like we described in \Cref{sec:flexibility}.

ECROs and Katara allow a programmer to influence a CRDT's semantics by providing application invariants in formal logic. ECROs uses these invariants to derive rules like ``an edit operation should be applied before a concurrent delete operation'' (enforcing a delete-wins semantics). At runtime, it attempts to apply concurrent operations in a sequential order respecting these rules; if cyclic dependencies prevent such an order, it arbitrarily \emph{discards} operations until the cycles disappear. These cycles will be numerous in any setting with long-lived concurrent sessions (e.g., offline work), causing many discarded operations. Katara uses program synthesis to find code respecting the application invariants. However, the synthesized code's overall semantics are opaque, and so far only synthesizes simple CRDTs (e.g., counters and sets).

\paragraph{CRDT Algorithms and Composition Techniques}
Collabs's built-in CRDT implementations use many specific algorithms from prior works, all cited in our documentation.

Several works describe or implement composition techniques for CRDTs. Collabs is heavily influenced by these works, and our CRDT implementations use compositional designs whenever possible.
Leijnse, Almeida, and Baquero~\cite{op_based_patterns} and Baquero et~al~\cite{state_based_patterns} describe classic op-based CRDTs in terms of compositional designs. Kleppmann and Beresford design a CRDT for JSON-formatted data that is motivated in terms of nested maps and arrays \cite{json_crdt}. Weidner, Miller, and Meiklejohn design a compositional construction for op-based CRDTs that they call the semidirect product of CRDTs \cite{semidirect}. We implement many of these techniques in Collabs.

Riak \cite{riak_datatypes}, Bloom$\mbox{}^L$ \cite{bloom_lattices}, and LVars \cite{lvars} implement CRDT-valued maps and ``stores'' that use several CRDTs side-by-side. Those are the basis for our \texttt{CObject} (\Cref{sec:system2}). Some of our CRDT-valued collections, including \texttt{CList}, are based on Yjs's CRDT-valued collections. Unlike those collaboration frameworks, Collabs allows not just specific composition techniques, but also arbitrary programmer-added composition techniques, due to dynamism (\Cref{sec:system3}).


\vspace{-5mm}
\section{Conclusion}
\label{sec:conclusion}
\vspace{-4mm}
We described Collabs, a CRDT-based collaboration framework that prioritizes both semantic flexibility and performance flexibility. By allowing programmers to implement their own CRDTs---either from scratch or by composing existing building blocks---Collabs enables nuanced, app-specific behaviors and performance optimizations. Thus Collabs functions as a CRDT laboratory, where researchers can try out new ideas, without compromising on performance or correctness.

In addition to generality, we showed that Collabs achieves superior performance on a many-user collaborative rich-text editing benchmarks. In particular, it scales to many more users than Google Docs, with lower end-to-end latency. 

\section*{Acknowledgments}
We thank Justine Sherry and Umut Acar for providing feedback on drafts of this paper, Kristof Jannes for providing OWebSync's eDesigners benchmark code, and Ignacio Musetti for implementing Collabs's minesweeper demo frontend. Matthew Weidner was supported by an NDSEG Fellowship sponsored by the US Office of Naval Research. Benito Geordie and Gregory Schare were supported by REUs sponsored by the US National Science Foundation. Yicheng Zhang was supported by a Cylab Secure and Private IoT Initiative Award.

\bibliographystyle{plain}
\bibliography{references}

\appendix

\section{Composition Examples}
\label{app:composition}

This appendix gives further examples of complex CRDTs defined through composition, expanding on \Cref{sec:composition}.

\subsection{Data Modeling: Ingredients}
\label{app:data_modeling}
\Cref{sec:composition} demonstrated \texttt{CRecipe}, a custom CRDT object modeling a recipe in our recipe demo. Our demo likewise models each ingredient as a \texttt{CIngredient}, which is also a CRDT object. Here is an outline of its class:
\begin{lstlisting}[language=TypeScript]
class CIngredient extends CObject {
  private _text: CText;
  private _amount: CScaleNum;
  private _units: CVar<Unit>;

  setUnits(units: Unit) {
    this._units.value = units;
  }
  ...
}
\end{lstlisting}

CRDT objects do require some extra boilerplate to set up the tree of Collabs. Specifically, you must assign a name to each instance field using the \texttt{CObject.registerCollab} method:

\begin{lstlisting}[language=TypeScript]
class CIngredient extends CObject {
  constructor(init: InitToken) {
    super(init);

    // First argument is the child's *name*, second argument calls its constructor.
    this._text = super.registerCollab("text", (textInit) => new CText(textInit));
    this._amount = super.registerCollab("amount", (amountInit) => new CScaleNum(amountInit));
    this._units = super.registerCollab("units", (unitsInit) => new CVar(unitsInit, Unit.GRAMS)); // Unit.GRAMS is initial value.
  }
}
\end{lstlisting}

\subsection{Composed Algorithms: List-with-Move}
\label{app:composed_algs}
In our recipe editor, recall that if one user moves an ingredient while another user edits the ingredient concurrently, both changes are preserved, as shown in \Cref{fig:move_correct}.

\begin{figure}
    \centering
    \includegraphics[width=\columnwidth]{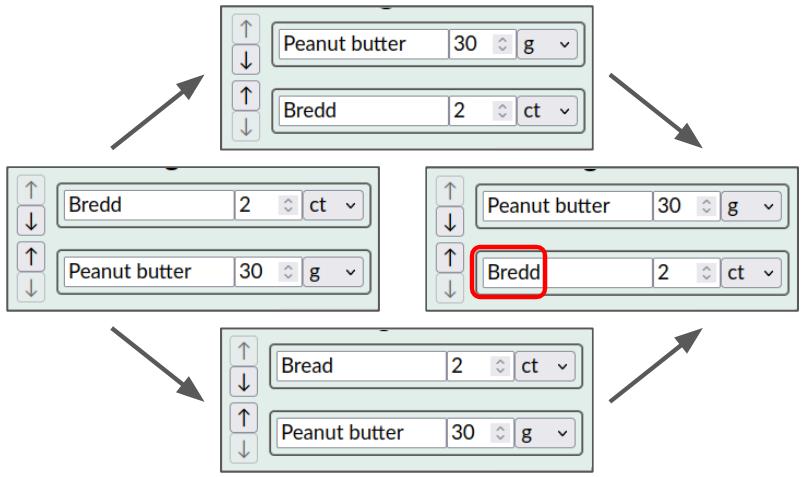}
    \caption{Potential anomalous behavior when one user moves an ingredient while another edits it concurrently: the typo fix for ``Bredd'' is lost.}
    \label{fig:move_wrong}
\end{figure}

Traditional list CRDTs assume that each element has an immutable position relative to other elements. Thus the only way to move an element is by cut-and-paste. However, if we move an ingredient in this way, then concurrent edits to that ingredient are lost, as \Cref{fig:move_wrong} shows.

Kleppmann describes a list-with-move CRDT that avoids this anomaly \cite{list_with_move}, but not as a traditional low-level CRDT algorithm. Instead, it is described as a composition of established CRDTs: a CRDT set of pairs (list element, current position). To move an element, you set its current position to a new value, which points into a separate list CRDT. 


Collabs lets us implement this compositional design directly, as part of the built-in \texttt{CList} CRDT. \Cref{fig:clist_state} shows a code scaffold. Each list element of type \texttt{C} is stored in a \texttt{CListEntry<C>}, which represents a pair $(\textrm{list element}, \textrm{current position})$. The list itself consists of:
\begin{itemize}
    \item A CRDT set (\texttt{CSet}) of \texttt{CListEntry}s, representing the algorithm's set of pairs. \texttt{CSet} manages the distributed initialization and deletion of pairs.
    \item A \texttt{CTotalOrder} that represents the separate list CRDT---specifically, its abstract total order on \texttt{Positions}.
    \item A \texttt{LocalList} that stores the list elements in order. This is not a CRDT itself but instead caches a local view of the list. Its state is a pure function of the previous fields.
\end{itemize}

\begin{figure}[t]
\begin{lstlisting}[language=TypeScript]
class CListEntry<C extends Collab> extends CObject {
  readonly value: C;
  // Position points into a separate list CRDT.
  readonly position: CVar<Position>;
}

export class CList<C extends Collab, /* ... */> {
  private readonly set: CSet<CListEntry<C>, /* ... */>;
  private readonly totalOrder: CTotalOrder;
  private readonly list: LocalList<C>;
}
\end{lstlisting}
\caption{Code scaffold for Collabs's implementation of the list-with-move.}
\label{fig:clist_state}
\end{figure}

\Cref{fig:clist_move} shows the \texttt{CList.move} method (slightly simplified), which moves an element from \texttt{startIndex} to \texttt{insertionIndex}.

\begin{figure}[t]
\begin{lstlisting}[language=TypeScript]
move(startIndex: number, insertionIndex: number) {
  // Position to insert at.
  const prevPos = this.list.getPosition(insertionIndex - 1);
  const nextPos = this.list.getPosition(insertionIndex);
  const position = this.totalOrder.createPosition(prevPos, nextPos);
  // CListEntry for the value to move.
  const entry = this.entryFromValue(this.list.get(i));
  // Move it.
  entry.position.value = position;
}
\end{lstlisting}
\caption{Collabs's \texttt{CList.move} implementation.}
\label{fig:clist_move}
\end{figure}

The complete \texttt{CList} implementation is 300 lines of code. Most of this code goes towards maintaining \texttt{CList.list} as a view of the CRDT state; technically, this is just an optimization. Thanks to the compositional design, we did not need to re-implement algorithms for dynamic CRDT creation (\texttt{CSet}, 270 lines) or list ordering (\texttt{CTotalOrder} and \texttt{LocalList}, 860 lines). Moreover, the result automatically satisfies the two guarantees from \Cref{sec:model}---strong convergence, and equivalent op- and state-based semantics---because the components do.

Object-oriented encapsulation ensures that the underlying components are hidden from \texttt{CList} users. In particular, the recipe editor app merely calls
\texttt{CList.move(startIndex, insertionIndex)}
to move an ingredient with the ``correct'' behavior from \Cref{fig:move_correct}.

This example demonstrates how Collabs's composition and encapsulation let us implement useful but nontrivial semantics, with a simple external API.

\section{Capabilities Evaluation: Explanations}
\label{app:capabilities}

\paragraph{Local-first}
The framework is local-first in the sense of Kleppmann et~al.~\cite{local_switch}: it allows local edits even when offline for an indefinite period of time, and it allows decentralized (server-optional) collaboration. Only certain CRDT-based frameworks satisfy this (Collabs, Yjs, Automerge, Legion).

The other frameworks allow offline edits but not decentralized collaboration. ShareDB uses centralized Operational Transformation and Replicache uses server-side reconciliation, both of which require a central server for collaboration. OWebSync uses CRDTs but also requires a central server.

\paragraph{Rich-text editing}
The framework has built-in support for collaborative rich-text editing. This is a popular but difficult use case, requiring a powerful collaboration framework. Only the libraries that we benchmark in \Cref{sec:rich_text_perf} have this capability.

\paragraph{Nested data}
The framework supports arbitrarily nested data, such as lists that can contain lists ($\ldots$) or JSON. This is necessary for complex, composed application state. Collabs and Yjs allow collections of CRDTs (which may themselves be collections, etc.), while Automerge, ShareDB, and OWebSync have JSON data structures.

\paragraph{List-with-move}
The framework supports moving elements in lists with the intuitive semantics shown in \Cref{fig:move_correct}. Only Collabs has a built-in implementation, as described in \Cref{app:composed_algs}.

Yjs and Automerge's authors have suggested workarounds in response to programmers' requests for this feature, but they are incomplete. Jahns sketches code that runs on top of a Yjs map; it does not work in all scenarios because it uses fractional indexing instead of a proper list CRDT \cite{move_yjs}. Kleppmann describes in prose how to use an Automerge map and list in concert to implement his original algorithm; that does not provide indexed access to the actual list because the internal list may contain duplicates \cite{move_automerge}.


\paragraph{Encapsulated data models}
The framework lets an app define encapsulated, type-safe data models for portions of its own state. We gave a Collabs example with \texttt{CRecipe} in \Cref{sec:composition}; \texttt{CIngredient} in \Cref{app:data_modeling} is another.

Jahns demonstrates a Yjs data model for collaborative Jupyter notebooks, which wraps Yjs maps and list CRDTs in a Jupyter-specific class \cite{yjs_jupyter_blog_post}. However, such data models are not first-class objects in Yjs: unlike built-in CRDTs, they cannot be used as values in CRDT collections.

The other frameworks that support nested data do so in the form of JSON. This does not permit encapsulation or other object-oriented programming techniques. At best, one can define reusable functions that operate on a given JSON path in a data-specific way.


\paragraph{Semantic flexibility \& performance flexibility}
The framework permits both semantic flexibility and performance flexibility as defined in the Introduction. We described Collabs's semantic flexibility in \Cref{sec:flexibility} and performance flexibility in \Cref{sec:system1}.

Replicache lets programmers express operations in the form of custom ``mutators''. For example, CRDT counter semantics are not built in to Replicache, but a programmer can implement them using a mutator that increments a value \cite{aboodman_lfw_talk}. 

ShareDB allows a programmer to supply a custom transformation function to its core Operational Transformation algorithm. In principle, this permits semantic flexibility, but designing correct transformation functions is known to be difficult \cite{ot_correctness}. Performance flexibility is limited because all operations still go through ShareDB's core algorithm.

We argue that Yjs and Automerge have limited flexibility even when you are allowed to modify their source code. All Yjs CRDTs must be wrappers around its core list CRDT; e.g., a counter must be modeled as an ever-growing list of increment operations. 
To add a new operation to Automerge, one must add support for it to several internal modules, each of which implements one feature (e.g., serialization) for every operation type. E.g., the pull request that added rich-text formatting modified over a dozen files concerning mutations, serialization, observers (events), queries, and transactions \cite{automerge_pr_rich_text}.


\section{Additional Benchmark Info}
\begin{figure}[h]
    \centering
    \includegraphics[width=0.9\columnwidth]{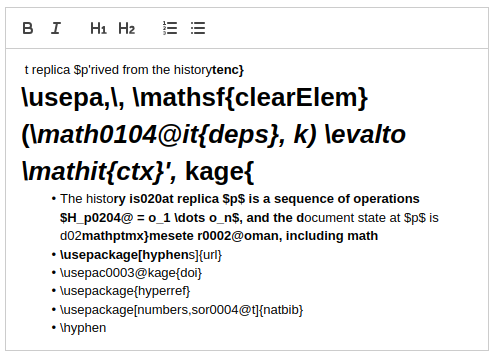}
    \caption{Screenshot of the benchmark's collaborative rich-text editor.}
    \label{fig:quill_screenshot}
\end{figure}

\Cref{fig:quill_screenshot} shows a screenshot of the collaborative rich-text editor GUI used for all experiments except Google Docs. It uses Quill with a basic set of allowed formats.

\begin{figure}[h]
    \centering
    \includegraphics[width=0.8\columnwidth]{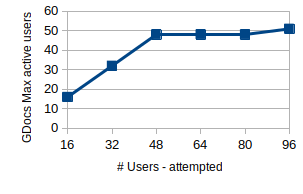}
    \caption{The maximum number of users who managed to type in the Google Doc, as a function of the number attempted.}
    \label{fig:gdocs_max}
\end{figure}

\Cref{fig:gdocs_max} shows the number of users able to connect to a Google Doc in our experiments. We attempted to open a Google Doc with 16--96 users, then measured the number of users who actually managed to type in the document (specifically, they observed at least one of their own sigils render).

\begin{table*}[ht!]
    \centering
    \renewcommand{\arraystretch}{1.1}
    \addtolength{\tabcolsep}{-0.2em}
    \small
    \begin{tabular}{lcccccc}
        Framework & \# Users & \shortstack{Saved state \\ size (KiB)} & \shortstack{Ratio to \\ Quill state size} & Load time (ms) & Save time (ms) & \shortstack{Memory \\ usage (MiB)} \\ \hline
        Collabs & 16 & 74 & 2.2 & $22.5 \pm 1.2$ & $2.7 \pm 0.2$ & $1.4 \pm 0.1$ \\
        CollabsNoVC & 16 & 75 & 2.2 & $21.4 \pm 0.6$ & $2.7 \pm 0.0$ & $1.4 \pm 0.1$ \\
        Yjs & 16 & 90 & 2.7 & $8.4 \pm 1.4$ & $3.7 \pm 0.1$ & $1.1 \pm 0.1$ \\
        Automerge & 16 & 265 & 5.9 & $458.8 \pm 4.3$ & $83.0 \pm 1.8$ & $61.8 \pm 2.2$ \\
        \hline
        Collabs & 112 & 557 & 2.3 & $327.5 \pm 4.6$ & $22.1 \pm 0.3$ & $10.5 \pm 0.1$ \\
        CollabsNoVC & 112 & 547 & 2.2 & $279.7 \pm 4.4$ & $22.0 \pm 0.4$ & $10.4 \pm 0.1$ \\
    \end{tabular}
    \caption{CRDT Metadata overhead for the saved state at the end of the given OptQuill experiments. For load/save times and memory usage, we report mean $\pm$ standard deviation for 10 trials, which followed 5 warmup trials.}
    \label{tab:metadata}

\end{table*}

\subsection{Metadata Overhead Data}
\label{app:metadata_perf}
We evaluated Collabs's metadata overhead (\Cref{sec:metadata_perf}) as follows.

At the end of each OptQuill rich-text editing experiment, we asked the collaboration framework to save a copy of its own state. (For the CRDT frameworks, this is a mergeable state-based CRDT state.) We then compared the state's size to the size of Quill's own representation of the state, which describes the text and formatting without collaborative metadata.\footnote{Due to nondeterminism and differences between frameworks, Quill's state size differs between experiment runs.} We also measured how long it takes to load and save the state, and the memory used by a saved state.\footnote{For Collabs (both variants) and Yjs, we measured the change in JavaScript's heap used when loading a document, running the garbage collector before each measurement. For Automerge, which uses WASM instead of the JavaScript heap, we instead measured the change in resident set size.} All experiments used Node.js v18.15.0 running on an Ubuntu 22.04.3 LTS with a 4-core Intel i7 CPU @1.90GHz and 16GB RAM.

\Cref{tab:metadata} shows the results for 16 users (where all frameworks were fully functional) and 112 users (the max where Collabs was fully functional). We see that Collabs has a small memory footprint and ratio to Quill state size, even for the large document at the end of the 112 user experiment (245 KiB Quill state size; 120 KiB of plain text). While loading the large document takes $\approx 300$ ms, this only needs to be done at the start of an app, and it is comparable to the network latency that a local-first app avoids.

Yjs also has practical performance for 16 users, while Automerge has relatively large load time and memory usage. We attribute this to Automerge's design choice to retain a full history of operations, in the style of a version control system. Collabs and Yjs instead deliberately trim or compress stale metadata (tombstones); this rules out some features but is more practical for ``live'' usage.

\end{document}